\documentclass[sigconf, nonacm]{acmart}
\usepackage{listings}
\usepackage{multirow}
\usepackage[table]{xcolor}
\usepackage{array} 
\usepackage{xcolor} 

\newfloat{sql}{htbp}{losql}
\floatname{sql}{Sql}

\begin{document}
\title{Efficient Cost-Based Rewrite in a Bottom-Up Optimizer}

\settopmatter{authorsperrow=4}

\author{Qi Cheng}
\affiliation{%
  \institution{Huawei, Cloud BU}
  \streetaddress{19 Allstate Pkwy, Markham}
  \city{Markham}
  \state{ON}
  \country{Canada}
  \postcode{L3R 5A4}
}
\email{qi.cheng1@h-partners.com}
\author{Yang Sun}
\affiliation{%
  \institution{Huawei, Cloud BU}
  \streetaddress{19 Allstate Pkwy, Markham}
  \city{Markham}
  \state{ON}
  \country{Canada}
}
\email{yang.sun5@huawei.com}
\author{Weidong Yu}
\affiliation{%
  \institution{Huawei, Cloud BU}
  \city{Markham}
  \state{ON}
  \country{Canada}
}
\email{weidong.yu@h-partners.com}
\author{Danny Chen}
\affiliation{%
  \institution{Huawei, Cloud BU}
  \city{Markham}
  \state{ON}
  \country{Canada}
}
\email{danny.chen@huawei.com}
\author{Weicheng Wang}
\affiliation{%
  \institution{Huawei, Cloud BU}
  \city{Markham}
  \state{ON}
  \country{Canada}
}
\email{ben.wang1@huawei.com}
\author{Chong Chen}
\affiliation{%
  \institution{Huawei, Cloud BU}
  \city{Markham}
  \state{ON}
  \country{Canada}
}
\email{chongchen@huawei.com }
\author{Per-\r{A}ke Larson}
\affiliation{%
  \institution{Huawei, Cloud BU}
  \city{Markham}
  \state{ON}
  \country{Canada}
}
\email{paul.larson@h-partners.com}

\begin{abstract}
The query optimizer in a Database Management Systems (DBMS), translates declarative queries into efficient execution plans. Conventional bottom-up optimization consists of two main stages: Query Rewrite (QRW) and Cost-Based Optimization (CBO). However, applying a rewrite rule during QRW may not always be beneficial; the best choice may depend on the (estimated) execution cost of the original and rewritten expressions. Fully exploiting such cost-dependent rules necessitates interleaving QRW with frequent CBO invocations, thereby incurring substantial overhead and often impractical optimization times. To mitigate this inefficiency, we introduce a novel cost-based rewrite framework for bottom-up optimizers. The core of our approach is a multi-level caching mechanism for intermediate CBO results aimed at eliminating redundant computation. Furthermore, we establish and exploit upper cost bounds to intelligently prune the search space during optimization. We also contribute methodological solutions for caching and reusing intermediate plan results within a bottom-up optimizer architecture. The framework has been implemented in the GaussDB optimizer. Experiments show that it significantly reduces overall optimization time, demonstrating the effectiveness of our approach.
\end{abstract}

\maketitle

\section{Introduction}
\label{sec:intro}

Modern data management systems are fundamental to the contemporary information technology landscape. As data volumes grow and business requirements become increasingly complex, the query optimizer within a Database Management System (DBMS) plays a pivotal role. By translating a declarative user query into an efficient physical execution plan, the query optimizer directly influences the overall throughput and responsiveness of the system.

A modern bottom-up query optimizer relies on two core components: Query Rewrite (QRW) and Cost-Based Optimizer (CBO). The QRW module applies logical transformation rules, such as predicate pushdown or subquery-to-join conversion, to restructure a query into a more efficient, equivalent form. Subsequently, the CBO module generates alternative physical execution plans, estimates their costs using a statistical model, and selects the most efficient one.

Query rewrites are driven by predefined rules. A subset of these rules, here called heuristic rewrite rules, 
are always applied, as they unconditionally improve efficiency. However, another class, here called cost-based rewrite rules, challenges this clean separation between the QRW and CBO phases. The applicability of these rules is not guaranteed; it depends 
on the database content.

Consequently, the optimizer cannot decide whether to apply such a rule during the QRW phase without consulting the CBO to evaluate the cost implications.

This architectural dilemma in bottom-up optimizers introduces significant computational overhead. For each candidate cost-based rewrite, the optimizer must pause the QRW process and invoke the CBO to obtain the estimated costs of the original and rewritten queries. When multiple such rules are considered, this repeated and interleaved invocation of the CBO dramatically increases optimization time, in some cases rendering it impractical.

To address this well-known efficiency challenge, we propose a novel framework for a bottom-up optimizer.
The framework implements a multi-level caching mechanism for intermediate CBO plan results at the base table, join, and subquery levels, eliminating redundant computations across multiple CBO invocations. To support this mechanism, we present a methodological solution for caching and reusing intermediate plan results within a bottom-up optimizer architecture. The framework also leverages an educated guess, derived from information available during the QRW phase, to quickly establish a tight cost upper bound. This bound intelligently prunes the search space, guiding the optimizer toward better plans earlier in the process. The proposed framework has been implemented in the GaussDB optimizer.

The rest of the paper is organized as follows: In \autoref{sec:related} we survey related works. \autoref{sec:gaussoptimizer} provides an overview of the GaussDB optimizer to establish context for the subsequent discussion. \autoref{sec:ourmethod} we describe our method in detail. \autoref{sec:analysis} demonstrates our results, and conclude with a discussion in \autoref{sec:futurework}.
 
\section{Related Work}
\label{sec:related}
\begin{figure}
  \centering
  \includegraphics[width=\linewidth]{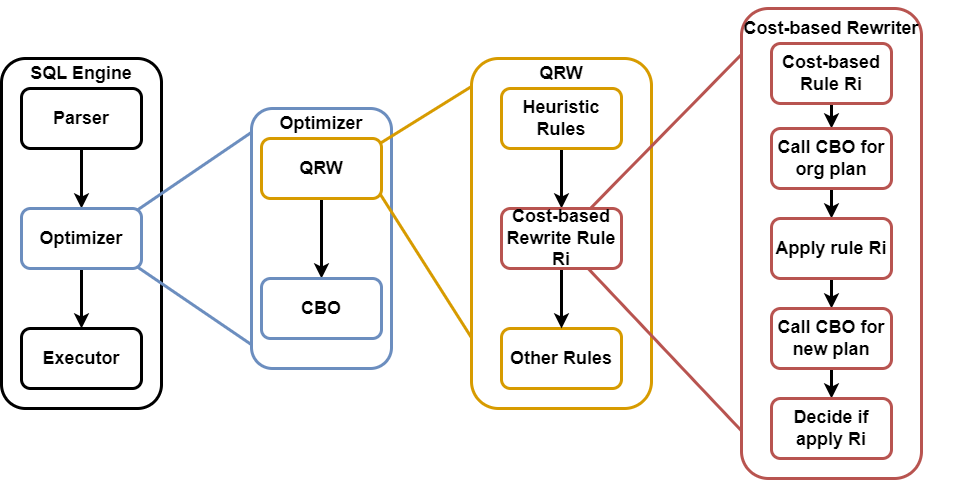}
  \caption{Na\"ive cost-based rewrite}
  \label{fig:fig1}
\end{figure}

As established, applying cost-based rewrite rules requires consulting the CBO to determine their benefit. A na\"ive implementation in a bottom-up optimizer simply applies the rules in a predefined order, invoking the CBO for each candidate rule. \autoref{fig:fig1} illustrates this process for a cost-based rewrite rule Ri.

The process in \autoref{fig:fig1} proceeds as follows:
\begin{enumerate}
    \item The QRW module takes the parsed SQL (Query Tree) as input and applies all heuristic rewrite rules prior to cost-based rewrite rule Ri, producing an intermediate query tree.
    \item QRW attempts to match this query tree against the pattern and conditions of cost-based rewrite rule Ri. If matching fails, Ri is skipped; otherwise, the process continues.
    \item Before applying Ri, QRW invokes the CBO to generate an execution plan and obtain its estimated cost for the current query tree.
    \item QRW applies rule Ri to produce a semantically equivalent but structurally different query tree.
    \item QRW invokes the CBO again to generate and cost an execution plan for the rewritten query tree.
    \item By comparing the costs from steps 3 and 5, QRW decides whether applying Ri is beneficial.
\end{enumerate}
This repeated invocation of CBO from QRW introduces substantial overhead, particularly when multiple cost-based rewrite rules are considered. Our analysis of the GaussDB optimizer reveals that the CBO accounts for more than 80\% of total compilation time for certain complex TPC-H queries, as shown in \autoref{table1}. Furthermore, enabling cost estimation within cost-based rewrite rules increases total compilation time by approximately 2.7 times for TPC-H queries (\autoref{tab:tpch_table} Column II), 2.6 times for TPC-DS queries (\autoref{tab:tpcds} Column II), and by a factor of 15 for a complex real-world customer query (\autoref{sql:sqlexample3})—from 13 ms to 205 ms.

\begin{table}
\caption{TPC-H Query Compile Time}
    \centering
    \begin{tabular}{|p{1cm}|p{1cm}|p{1cm}|p{1cm}|p{1cm}|p{1cm}|}
        \hline
        \textbf{TPC-H Query} & \textbf{Total Compile Time} & \textbf{QRW Run Time} & \textbf{QRW Percentage} & \textbf{CBO Run Time} & \textbf{CBO Percentage} \\
        \hline
        Q2 & 7.266 & 0.210 & 2.89\% & 6.007 & 82.67\% \\
        Q7 & 12.073 & 0.209 & 1.73\% & 9.870 & 81.75\% \\
        Q8 & 16.235 & 0.255 & 1.57\% & 14.259 & 87.83\% \\
        Q9 & 66.226 & 0.219 & 0.33\% & 63.246 & 95.50\% \\
        \hline
    \end{tabular}
    \label{table1}
\end{table}

Several research projects and commercial systems have attempted to mitigate this overhead. Oracle employs a block-level caching mechanism ~\cite{Ahmed06} that enables reuse of optimization results for unchanged query blocks. This approach effectively reduces redundant computations for repeatedly referenced subqueries. However, if a rewrite rule modifies a query block, it must be reoptimized from scratch.

Polar DB introduces a lightweight CBO variant called mini-CBO ~\cite{baba}. When cost-based rewrite rules require cost estimation, this lightweight optimizer is invoked instead of the full CBO to reduce overhead. The effectiveness of this approach depends critically on the accuracy of mini-CBO; if its cost estimates diverge from those of the full CBO, it may mislead QRW decisions.

DB2 adopts a fundamentally different approach ~\cite{Pirahesh92}, maintaining a strict separation between QRW and CBO by prohibiting cost-based rewrite rules entirely. All rewrite rules must guarantee semantic preservation and unconditional performance improvement. While this approach eliminates the overhead of repeated CBO invocations, it sacrifices potential optimization opportunities that require cost-based evaluation.

The existing literature on this area has predominantly focused on top-down optimizers employing a Cascades-style framework ~\cite{Graefe1995Cascades, Graefe1991Volcano, Chaudhuri99, Soliman2014Orca}. These systems, exemplified by implementations such as Orca and SQL Server, utilize a memo structure to store the results of transformation rules, including query rewrite rules and implementation rules. By organizing logically and physically equivalent expressions into groups within the memo, and by performing optimization in a single optimization pass, top-down optimizers inherently facilitate the caching of intermediate results, thereby avoiding redundant computations of semantically identical subexpressions.
However, to avoid blowing up optimization time, they cannot blindly apply all valid rewrite rules, instead which rules to apply is 
decided based on easy-to-compute heuristics.

In contrast, research addressing cost-based rewrite efficiency in bottom-up optimizers remains limited, with only a few studies exploring reuse mechanisms at the subquery level ~\cite{Zhu05, Meduri11}. Unlike their top-down counterparts, bottom-up optimizers do not retain alternative query trees for each rewrite rule. Furthermore, they typically invoke the cost-based optimizer (CBO) in separate optimization passes—both before and after applying cost-based rewrite rules. This separation complicates the reuse of intermediate results, as such results are generated and evaluated across distinct optimization passes, presenting a significant challenge for efficient plan caching and reuse, thereby demanding innovative methodological solutions.

Recent advances in large language models (LLMs) and machine learning have introduced novel paradigms for query optimization, particularly in the area of query rewrite ~\cite{Song25, Sun2024RBotCoRR, Wang2022WeTune, Li2024LLMR2CoRR, Dharwada2026LITHE, Zhou2021Learned, Liu2024Query}. The work most closely related to our own are Learned Rewrite ~\cite{Zhou2021Learned} and LLM-R\textsuperscript{2} ~\cite{Li2024LLMR2CoRR}. Learned Rewrite employs a Monte Carlo Tree Search to optimize the selection of applicable rewrite rules, while LLM-R\textsuperscript{2} leverages an LLM to recommend rule sets for a database rewrite system. However, both researches focus primarily on identifying an optimal set of rules to apply to a given query, rather than on improving the efficiency of applying individual cost-based rewrite rules. Although we are not aware of existing research that utilizes pre-trained LLMs to enhance the efficiency of cost-based rule application, we contend that AI-assisted methodologies represent a significant and complementary direction for the future development of query optimizers.

\section{GaussDB Optimizer}
\label{sec:gaussoptimizer}
\begin{figure}
  \centering
  \includegraphics[width=\linewidth]{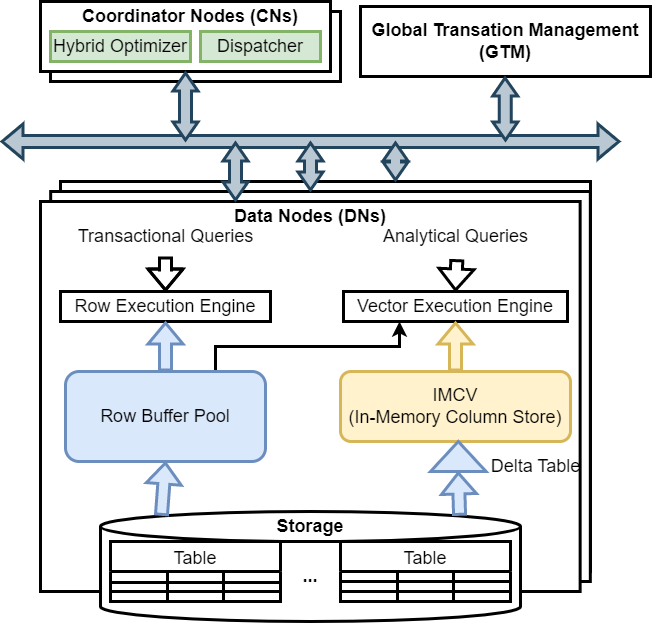}
  \caption{GaussDB Architecture}
  \label{fig:fig2}
\end{figure}

\begin{figure}
  \centering
  \includegraphics[width=0.8\linewidth]{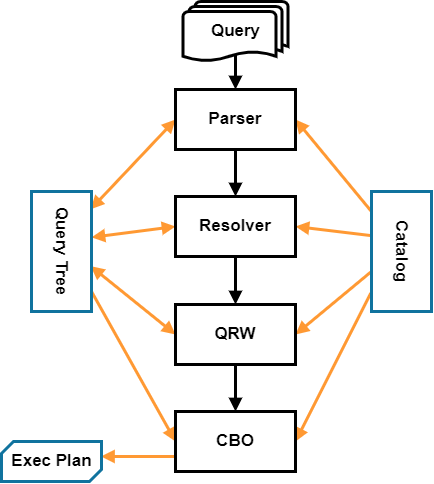}
  \caption{GaussDB Optimizer}
  \label{fig:fig3}
\end{figure}

GaussDB ~\cite{GaussDB2020, Memarzia2024GaussDB} is a cloud-native, distributed relational database system built on a shared-nothing architecture. As illustrated in \autoref{fig:fig2}, it employs a classic coordinator-node (CN) and data-node (DN) model: CN nodes act as the SQL entry point, responsible for parsing, optimization, and global execution planning, while DN nodes store data shards and execute query fragments in parallel. The optimizer operates in two distinct phases—Query Rewrite (QRW) and Cost-Based Optimization (CBO), as depicted in \autoref{fig:fig3}.

Upon receiving a query, the system first parses it and constructs an internal representation known as a query tree. The resolver then maps the data objects in the query tree to internal catalog objects. In the QRW phase, predefined rewrite rules are applied to transform the query tree into a semantically equivalent but more efficient form. Subsequently, the CBO phase devises an optimal execution strategy.

The GaussDB optimizer, as is typical for bottom-up optimizers, does not preserve multiple intermediate query trees during rule application; only a single "winner"  is retained. For a cost-based rewrite rule, the QRW engine directly invokes the CBO to estimate the costs before and after the rewrite and to determine whether the rule should be applied, as illustrated in \autoref{fig:fig1}. Each CBO invocation allocates temporary data structures to hold intermediate results, which are subsequently destroyed at the end of that optimization pass. To improve efficiency, these structures are often compressed using bit-set encodings of data objects—an encoding scheme that remains valid only within a single query block and optimization pass. At present, the GaussDB optimizer incorporates only seven cost-based rewrite rules. While the inclusion of additional rules is planned, their implementation is currently constrained by the potential increase in compilation time.

\section{Our Method}
\label{sec:ourmethod}
The na\"ive approach depicted in \autoref{fig:fig1} suffers from two primary inefficiencies in a bottom-up optimizer:
\begin{enumerate}
    \item It regenerates execution plans from scratch for each cost-based rewrite rule evaluation.
    \item It lacks a tight cost upper bound to effectively prune the search space during planning.
\end{enumerate}
While existing research and commercial systems have attempted to improve the efficiency of cost-based rewrites, to the best of our knowledge, no solution in a bottom-up optimizer comprehensively addresses both issues. The potential of reusing intermediate planning results remains underexplored, and the strategic use of a cost upper bound has been largely overlooked. This paper proposes a framework specifically designed to resolve these shortcomings in bottom-up optimizers.
Our approach is grounded in two key observations:
\begin{enumerate}
    \item Whether a rule will be beneficial can often be predicted based on   
    query characteristics and statistics, such as query structure, predicates, join types, and table statistics, which are available during the QRW phase and without invoking the CBO.
    \item Intermediate planning results, including base table accesses, joins, and untouched subqueries, are often reusable across multiple CBO invocations for the same query, as much of the underlying query structure remains consistent.
\end{enumerate}
These observations inform our proposed framework, which is designed to: (1) leverage information available during QRW to guide rule application, (2) maximize the reuse of intermediate CBO results to eliminate redundant computation, and (3) attempt to prune out unpromising results early.

During the QRW phase, when evaluating a cost-based rewrite rule, our method employs an "educated guess" based on available statistics, table physical properties, system resources, and query patterns to predict whether applying the rule will be beneficial. This guess is fast, as it bypasses the cost-based optimizer (CBO) and thus avoids the combinatorial exploration of access paths and join strategies, and is likely accurate when based on a deep understanding of the rule's benefit scenarios. The predicted "winning" query tree is then passed to the CBO. The resulting plan cost serves as a tight upper bound, which is used to prune more expensive paths early during subsequent planning. This upper bound is continuously updated as lower-cost plans are found.

In time-critical scenarios, such as OLTP, the predicted winner can be applied directly without cost comparison, foregoing the full CBO evaluation for the alternative plan.

When invoking CBO to calculate the cost of both the original and rewritten query trees, our framework caches plans for each query block as well as intermediate planning results. If a query block is unmodified by a rewrite rule, its cached plan is reused. For altered structures, the system maximizes reuse by leveraging cached intermediate planning results at multiple levels of granularity. While intermediate plan caching (memoization) is standard in top-down optimizers ~\cite{Graefe1995Cascades, Chaudhuri99, Ding24}, our work is, to the best of our knowledge, the first to introduce and adapt it effectively in a bottom-up optimizer.

It is important to note that our method only accelerates the determination of a cost-based rewrite rule's applicability. This process is orthogonal to the broader application order of rules within the query rewrite engine. Once applicability is decided, the engine proceeds according to its predefined strategy—whether that involves applying subsequent rules to the new query tree, re-applying the same rule to different sub-trees, or restarting the entire rule set.

\subsection{Educated Guess}
\label{sec:edguess}
As a core component of our framework, we introduce an "Educated Guess" mechanism for each cost-based rewrite rule within the QRW phase. For each cost-based rewrite rule, we predefine decision logic that leverages information available during QRW, including statistics (e.g., base table statistics, estimated cardinalities), table physical properties (e.g., indexes, partitioning), query patterns, and system resources, to quickly determine whether applying the rule is likely to be beneficial. This heuristic predicts which alternative is expected to be faster without invoking the expensive CBO.

It bears emphasizing that the Cost-Based Optimizer (CBO) retains the final authority and still evaluates both scenarios—applying the rewrite rule versus not applying it. The educated-guess heuristic only prioritizes the more promising alternatives first. This initial cost, which is anticipated to be lower, establishes a tight upper bound. This bound can then enable early pruning during the subsequent evaluation of the less promising alternative, thereby optimizing the overall decision process. It is important to note that it is perfectly fine for a rule to rely on a default prediction, for example, always assume that the rewritten query tree will be the winner. A wrong prediction may slow down optimization, but it won't change the final result.

Since the educated-guess logic is inherently rule-specific, a uniform algorithm for all cost-based rewrites is impractical. Instead, we demonstrate its application through two concrete examples, each exemplifying a typical category of cost-based rewrite rules.

\subsubsection{Example: Subquery merge}
\label{sec:examplemerge}
The following example demonstrates the educated guess mechanism for a rule that merges a subquery into its parent query. To preserve the original semantics, a $DISTINCT$ operation is added, shown in \autoref{sql:sqlexample1}.

\begin{sql}
\begin{lstlisting}
-- Original Query
SELECT T1.pk, T1.c1, T1.c2 FROM T1 WHERE T1.c1 IN 
(SELECT c1 FROM T2 WHERE T2.c2 <= T1.c2 AND T2.c3 = 10);
-- Rewritten Query
SELECT DISTINCT T1.pk, T1.c1, T1.c2 FROM T1, T2 
WHERE T1.c1 = T2.c1 AND T2.c2 <= T1.c2 AND T2.c3 = 10;
\end{lstlisting}
\caption{Subquery merge}
\label{sql:sqlexample1}
\end{sql}

While this rewrite enables more join order and method possibilities, often leading to significant performance gains ~\cite{Pirahesh97}, it introduces a potentially expensive $DISTINCT$ operation that may degrade performance if applied to a large intermediate result, especially it spills to disk. Our educated guess for this rewrite rule uses the statistics of \lstinline{T1} and \lstinline{T2}, estimates the join result between \lstinline{T1} and \lstinline{T2}, and checks the available memory resource to decide whether to apply the rewrite rule. This demonstrates how QRW can make informed decisions independently of the CBO, using readily available information to predict whether a rewrite is more likely to lead to an improved plan.

\subsubsection{Example 2: Eager aggregation}
\label{secexampleeger}
The following example demonstrates our educated guess approach for an Eager Aggregation ~\cite{Yan95} rewrite rule in TPC-H Q10.  Under the conditions that $c\_custkey$, $n\_nationkey$, and $o\_orderkey$ are primary keys, the eager aggregation can be simplified as shown in \autoref{sql:sqlexample2}.

\begin{sql}
\begin{lstlisting}
-- Original Query
SELECT c_custkey, c_name, sum(l_extendedprice * (1 - l_discount)) as revenue, c_acctbal, n_name, c_address, c_phone, c_comment 
FROM customer, orders, lineitem, nation 
WHERE c_custkey = o_custkey AND l_orderkey = o_orderkey AND o_orderdate >= date '1993-10-01' AND o_orderdate < date '1993-10-01' + interval '3' month  AND l_returnflag = 'R' AND c_nationkey = n_nationkey 
GROUP BY c_custkey, c_name, c_acctbal, c_phone, n_name, c_address, c_comment 
ORDER BY revenue DESC 
LIMIT 20;
-- Rewritten Query
SELECT c_custkey, c_name, temp.revenue, c_acctbal, n_name, c_address, c_phone, c_comment 
FROM customer, nation, 
    (SELECT o_custkey, sum(l_extendedprice * (1 - l_discount)) as revenue 
    FROM orders, lineitem 
    WHERE l_orderkey = o_orderkey AND o_orderdate >= date '1993-10-01' AND o_orderdate < date '1993-10-01' + interval '3' month 
    GROUP BY o_custkey) as temp(o_custkey, revenue) 
WHERE c_custkey = temp.o_custkey AND c_nationkey = n_nationkey 
ORDER BY revenue DESC 
LIMIT 20;
\end{lstlisting}
\caption{Eager aggregation}
\label{sql:sqlexample2}
\end{sql}

While pushing the $GROUP\: BY$ operation down before joins can improve performance through early data aggregation, this strategy is only beneficial if the cost of the pushed-down aggregation is not prohibitively high. A critical factor in this cost is the selectivity of the subsequent join; if the join between the $nation-customer$ and $orders-lineitem$ tables filters out too many rows, the overhead of early aggregation is not justified. To address this, our educated guess leverages a HyperLogLog (HLL) based algorithm ~\cite{Flajolet07} to determine if the join is "near-lossless", i.e. if it preserves most input rows, before proceeding with the push-down of the $GROUP\: BY$ clause. The $ORDER\: BY$ and $LIMIT$ operations are retained post-join to ensure correct results. The HLL structure is computed when gathering table statistics and subsequently used to estimate near-lossless-ness during optimization without requiring recomputation.
                                 
\begin{figure}
  \centering
  \includegraphics[width=\linewidth]{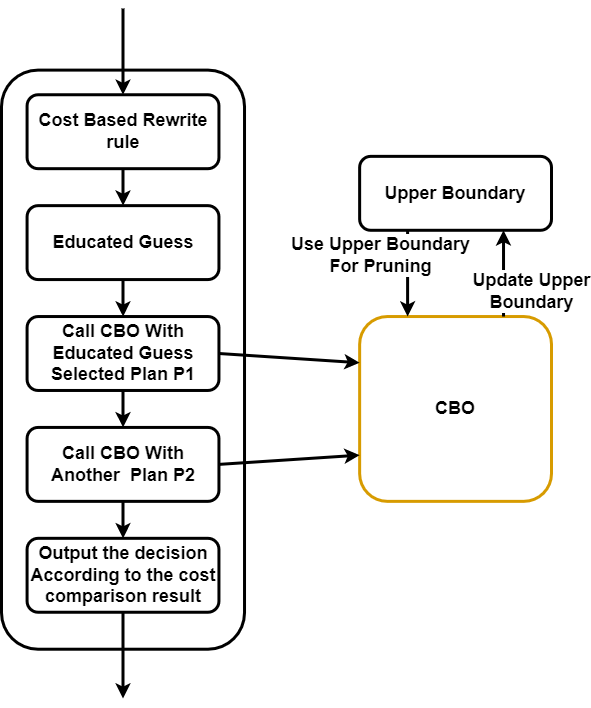}
  \caption{Educate Guess}
  \label{fig:figold2}
\end{figure}

As illustrated in \autoref{fig:figold2}, the Educated Guess mechanism yields a tight cost upper bound by directing the CBO to compute the plan cost for the most promising query variant first. This initial cost serves as a reference point that can be further refined as additional cost-based rewrite rules are evaluated. Given that transformations like subquery merging can improve query performance by orders of magnitude ~\cite{Pirahesh92}, this approach frequently yields a robust upper bound early in the optimization process. The bound then enables aggressive pruning of the search space and may permit early termination of planning for alternative query forms that cannot produce cheaper plans, thereby significantly reducing the search space. It is worth emphasizing that the "educated guess" does not require a high degree of precision, as the final decision is ultimately governed by the cost model. For cost-based rewrite rules that are beneficial in most cases but detrimental in specific instances—such as subquery merge—the heuristic logic can adopt the rewrite as a default behavior, while leveraging statistics, table physical properties, and system resource information to preclude application in negative outlier scenarios. For rules that are advantageous only in particular contexts—such as group-by pushdown—the heuristic can instead utilize the same data sources to explicitly identify and target those positive cases. In certain instances, it may even be acceptable to apply the rewrite as a default heuristic without further qualification.

\subsection{Query Plan Intermediate Result Reuse}
\label{sec:queryplaninterresultreuse}
The repeated invocation of the Cost-Based Optimizer (CBO) in the na\"ive method presents a clear opportunity to eliminate redundant work by reusing cost computations at multiple levels of granularity and  avoid redundant plan generation and cost computation work. In contrast to reuse only at the \textbf{query-block level}, we propose a framework that enables reuse across \textbf{multiple levels}:
\begin{itemize}
    \item \textbf{Base Table level}: For each table, the optimizer considers possible access methods based on statistics, relevant predicates, and physical characteristics (e.g., available indexes and data layout). By caching and reusing the resulting access paths and their cost components, we reduce the overhead of repeatedly evaluating access alternatives and cost estimation.
    \item \textbf{Join level}: For a joined table with specified join conditions, we cache the corresponding best join strategies and its computed costs. Subsequent encounters can directly retrieve these cached results if they can be reused, thereby avoiding expensive join planning.
    \item \textbf{Query-block level}: When query blocks identical to previously blocks without changes by QRW, we reuse the prior planning results for them to avoid repeated computations.
\end{itemize}

This multi-level caching and reuse framework fully exploits the potential of plan reuse in bottom-up optimizers, enabling efficient computation sharing throughout the optimization process. While plan caching and reuse are well-established in top-down optimizers, as discussed in \autoref{sec:related}, their application in bottom-up optimizers requires innovative solutions. We describe our method in detail in the following sections.

\subsubsection{Hash table and prefilter}
\label{sec:hashtabelprefilter}
In a bottom-up optimizer, the costs associated with a query before and after a cost-based rewrite are computed in two distinct optimization passes. To enable the reuse of intermediate results across these separate CBO invocations, the results from the first optimization pass must be cached and made retrievable during the second pass. To facilitate effective plan reuse, we first construct a global hash table to store the intermediate plan structures. These structures, which encapsulate various access paths for a single table, join strategies and plans for a group of tables, or complete plans for a query block, are central to the optimizer's search process. For context, these are analogous to the $RelOptInfo$ structure for tables and joins, and the $SubPlan$ structure for query blocks in PostgreSQL. 

The hash key, which we term the \textit{signature}, is an efficiently computed digest of key structural properties of the intermediate plan ~\cite{Zhou2007Efficient}. In our implementation, this signature includes the table OIDs, the count of Bloom filter candidates, the number of restrictions, the number of join predicates, the number of lateral relation tables, the number of output columns, the number of column-equivalence-class, and the number of subqueries. 

Before each intermediate step is planned by CBO, its signature is extracted and hashed. A mismatch in the signature indicates that the cached intermediate plan is not a match, allowing us to immediately reduce the search space of the cached plans with minimal overhead. Our analysis of TPCH queries, presented in \autoref{table2},  demonstrates that the signature method typically limits the search to a single cached plan in the majority of instances.
\begin{table}
\caption{TPCH Query Plan Cache Analysis}
    \centering
    \begin{tabular}{|p{1.4cm}|p{1.6cm}|p{1.2cm}|p{1.2cm}|p{1.2cm}|}
        \hline
        \textbf{Cached Plan Type} & \textbf{Total Bucket Visits} & \textbf{Bucket Length 1} & \textbf{Bucket Length 2} & \textbf{Bucket Length 3} \\
        \hline
        Base Table & 105 & 73 & 22 & 10 \\
        Join Table & 138 & 138 &  &  \\
        Subquery & 12 & 12 &  &  \\
        \hline
        \textbf{Total} & 255 & 223 & 22 & 10  \\
        \hline
    \end{tabular}
    \label{table2}
\end{table}

Although the signature can significantly reduce the search space for cached plans, the matched signature may still link to a list of cached intermediate plans since the hash key just extracts some basic signature information, but not all information of the intermediate plan structure. We need a full matching algorithm that can verify and select the correct intermediate plan from all the potential matches we find.

\subsubsection{Matching Intermediate Plan Structures}
\label{sec:matchingstructure}
A key challenge in matching intermediate plan structures lies in the table and column identifiers used within query plans. Plans typically reference tables through indices in a local array rather than by direct object IDs. However, this array is specific to each query block. The indices might become invalid when rewrites alter the query structure, such as removing join tables or merging subqueries. To enable plan matching, we establish mappings of the table and column identifiers between the pair of the plan structures and translate corresponding predicates and expressions accordingly. This translation allows us to effectively identify equivalent intermediate plans for reuse.

Our matching algorithm attempts to match intermediate plans synchronously as it builds the execution plan in a bottom-up manner. 
The plan matching process begins by establishing a mapping between table identifiers. All base tables in the local table arrays are matched by their OIDs, creating a mapping of table indices between the two plans.

\begin{figure}
  \centering
  \includegraphics[width=\linewidth]{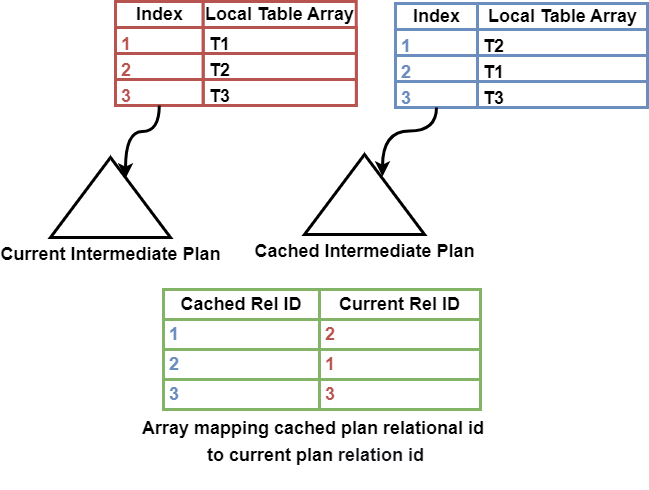}
  \caption{Base table relation id mapping}
  \label{fig:oldfig3}
\end{figure}

\autoref{fig:oldfig3} illustrates the mapping of base table identifiers between intermediate plan structures. The diagram depicts a pair of such structures: the cached plan (right) and the plan currently under construction (left). Each structure maintains a local table array, indexed by the red and blue numerals, which represent the tables \lstinline{T1}, \lstinline{T2}, and \lstinline{T3} from the query. The green rectangle highlights the logical mapping of relation identifiers from the cached plan to the current plan. For instance, relation ID 2 in the cached structure is mapped to relation ID 1 in the structure under construction.
Before the CBO plans base table access methods, we compute a signature to search the global cache. On a successful hash bucket match, we retrieve candidate intermediate plans. For each candidate, we use the established table mapping to translate predicates, expressions and all constructs that reference relation IDs (e.g., bit-sets encoding groups of relation IDs) into the current query context. Following translation, we perform a detailed structural comparison of all plan components—including Bloom filters, restrictions, output columns, and other metadata.

\begin{figure}
  \centering
  \includegraphics[width=0.8\linewidth]{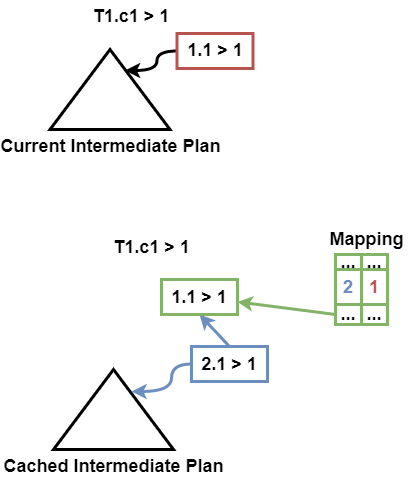}
  \caption{For predicate and expression translation}
  \label{fig:fig4}
\end{figure}

\autoref{fig:fig4} shows this translation process: a predicate \lstinline{T1.c1 > 1} from a cached plan (expressed as \lstinline{2.1 > 1}, where 2 is \lstinline{T1}'s table index, and 1 is the column ID) is translated to the current context, where \lstinline{T1}'s relation index is 1 using the table identifier mapping, resulting in \lstinline{1.1 > 1}.
The same approach extends to joined tables. Before planning join strategies, we extract a join signature for cache lookup. The subsequent matching process mirrors base table access: candidate retrieval, predicate, expression and relation-ID-related item translation, and structural verification. Furthermore, the algorithm supports self-joins, which occur when a single table is referenced multiple times in a query block; however, a full treatment of this feature is omitted for brevity. 
Subquery matching requires special handling. For untouched query blocks, we extract signatures and match them similarly to base tables and joins. However, successful matching requires comparing the entire query tree and establishing mappings for the subquery's output table and column IDs, which are then utilized during parent query block matching.

\begin{figure*}
  \centering
  \includegraphics[width=0.9\linewidth]{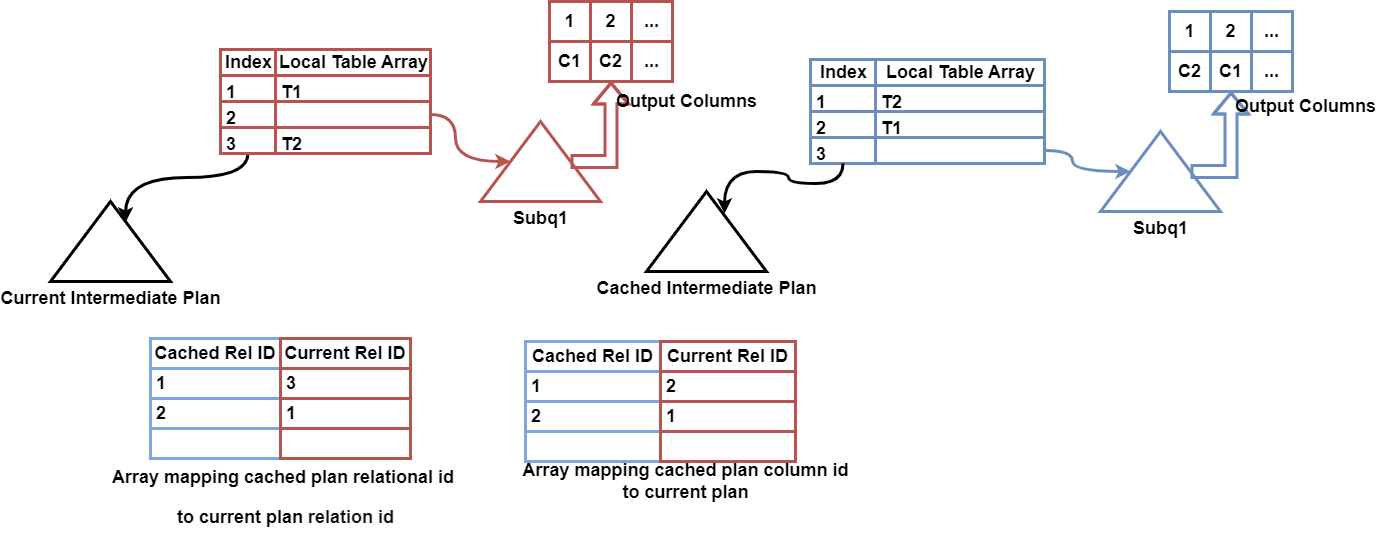}
  \caption{Subquery matching and output mapping}
  \label{fig:fig5}
\end{figure*}

\autoref{fig:fig5} demonstrates the process of subquery matching and output mapping. A pair of logically equivalent subqueries (Subq1) are identified within the cached and current planned plans. To facilitate plan reuse, the algorithm establishes a mapping for the subqueries' local table indices during the planning of their parent query blocks (see mapping table, left). Additionally, a mapping between their output columns is created to resolve any discrepancies in the column ordering. This comprehensive matching and mapping strategy allows the algorithm to traverse and reconcile multiple levels of the query block hierarchy in a bottom-up manner.

Even with successful signature matching and structural alignment, cached plan reuse is not guaranteed. The following sections detail exceptional cases requiring additional validation beyond these fundamental algorithms.

\subsubsection{Bloom Filters}
\label{sec:bloomfilter}
Our recent work has included the Bloom filter planning into Gaussdb optimizer ~\cite{Zeyl25}. Our two-phase Bloom filter optimization strategy first identifies all viable Bloom filter candidates applicable to individual tables by analyzing various logical join patterns. These candidates are then evaluated and incorporated during the physical planning process in the second phase. Because the first phase involves no physical planning, the reuse of intermediate plans is utilized exclusively in the second phase.  

\begin{figure}
  \centering
  \includegraphics[width=0.9\linewidth]{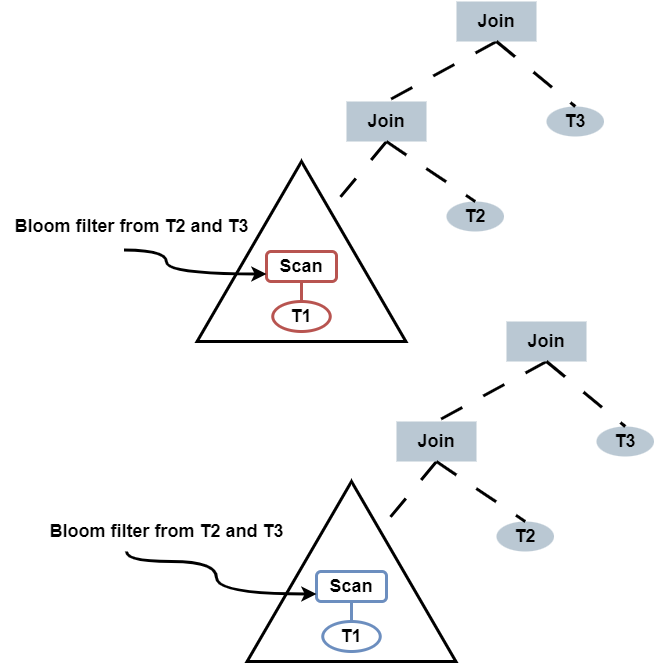}
  \caption{Bloom filter from other joins}
  \label{fig:fig6}
\end{figure}

When the Cost-Based Optimizer (CBO) plans access methods for a table, it must consider not only restriction predicates but also any applicable Bloom filters. Consider the example query illustrated in \autoref{fig:fig6}. The upper portion depicts the target plan to be constructed, while the lower portion shows a cached intermediate plan. In both plans, tables \lstinline{T1}, \lstinline{T2}, and \lstinline{T3} are joined, with filters derived from \lstinline{T2} and \lstinline{T3} applied to \lstinline{T1}. To successfully reuse a cached intermediate plan for accessing \lstinline{T1}, the optimizer must match not only the core properties (as defined in \autoref{sec:matchingstructure}) but also the specific set of Bloom filters applied to the table

\subsubsection{Join Orders}
\label{sec:joinorder}
The presence of outer, semi, or anti joins in a query constrains the valid join orders due to their semantic rules. Optimizers typically manage these constraints using specialized data structures. For example, PostgreSQL uses $SpecialJoinInfo$ to specify, for a group of tables, the minimum sets that must appear on the leftmost and rightmost sides of a joins. This structure guides the optimizer as it explores different join patterns

Reusing a cached plan for the join of a group tables requires more than just matching the involved tables and predicates. The data structures that guide join ordering must also be considered. However, these structures (e.g., $SpecialJoinInfo$) are typically defined at the query level, not for a specific join group. Therefore, to reuse a cached plan, the optimizer must first identify all query-level constraints that are relevant to the current join. If the join group's tables intersect with the leftmost or rightmost table sets defined by these constraints, then those constraints must be perfectly matched for the cached plan to be valid.

\subsubsection{Hints}
\label{sec:hints}
Optimizer hints represent another example of external guidance that influences optimizer behaviour through separate data structures. When a query includes hints mandating specific plan components, such as a join algorithm, the system must perform additional verification beyond matching the core properties. For instance, if a query specifies a Nested Loop Join for \lstinline{T1 JOIN T2}, any cached plan for this join must be validated to ensure it uses the required algorithm, as a differently-optimized plan would conflict with the hint constraints.

\subsubsection{In List to Join}
\label{sec:inlisttojoin}
\begin{figure}
  \centering
  \includegraphics[width=0.8\linewidth]{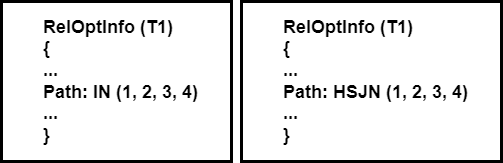}
  \caption{GaussDB RelOptInfo for in-list}
  \label{fig:fig7}
\end{figure}

When processing a query such as "\lstinline{SELECT * FROM T1 WHERE T1.c1 IN (1, 2, 3, 4);}", the optimizer typically converts the IN-list predicate into an equivalent join operation. Although this IN-list-to-join conversion is fundamentally a cost-based rewrite, many database systems, including PostgreSQL, GaussDB, and DB2, implement it within the Cost-Based Optimizer (CBO) rather than in the Query Rewrite (QRW). The optimizer generates alternative access paths for the table being filtered by the IN-list predicate. In some implementations, particularly in GaussDB, these alternative paths may result in non-standard plan structures, as illustrated in \autoref{fig:fig7}

The standard base table matching logic is insufficient here, as both plan structures ultimately reference the same base table (\lstinline{T1}) and have same restriction predicate. To address this, the system must first determine whether a cached intermediate plan result remains valid for the IN-list to Join transformation. Subsequently, during cache matching, it must guarantee retrieval of the specific variant that incorporates the specialized alternative access path.

\subsubsection{Column Equivalence Class}
\label{sec:coleqclass}
\begin{figure}
  \centering
  \includegraphics[width=0.8\linewidth]{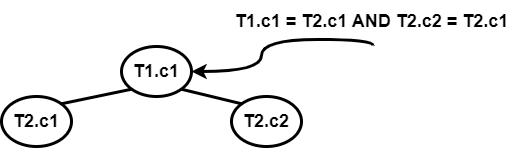}
  \caption{Column equivalence class}
  \label{fig:fig8}
\end{figure}

In many optimizers, equivalence predicates between columns are represented using Column Equivalence Classes (CECs). A CEC can be modeled as a minimum support tree of the connected component in the column equivalence graph. The nodes of this tree are the columns involved in the equivalence predicates, while the edges represent the individual equivalence relations that define the class. \autoref{fig:fig8} illustrates an example of such a column equivalence class.

Since semantically identical conditions can be represented with arbitrary internal sequences, direct list comparison often produces false mismatches. To compare two Column Equivalence Classes (CECs) correctly and efficiently for cache reuse, we avoid enumerating all possible permutations. Instead, we compare a canonical representation derived from the member set, such as a hash set of the members, ensuring a match based purely on the semantic set of equivalent attributes, independent of their storage order.

\section{Experimental Analysis}
\label{sec:analysis}
We evaluated our method using both TPC-H and TPC-DS benchmarks. For TPC-H, we used a TPC-H dataset with a scale factor of 10 (approximately 10 GB). Each of the 22 TPC-H queries was executed 6 times, and we report the average of the final 3 runs to reflect steady-state performance after data was loaded into memory. The dataset tables were stored in a row-oriented format. All experiments were conducted on an x86 server with 8 CPUs and 16 GB of memory.
\begin{table}
    \centering
    \caption{TPC-H Result}
    \label{tab:tpch_table}
    \rowcolors{2}{gray!15}{white}
    \begin{tabular}{lrrrrrr}
        \hline
        No. &I&II&III&IV&V&VI \\
        \hline
        Q2 & 2.836 & 7.609 & 6.186 & 29.81\% & 6.004 & 33.62\% \\
        Q4 & 1.053 & 1.810 & 1.623 & 24.76\% & 1.627 & 24.23\% \\
        Q7 & 4.958 & 12.545 & 10.190 & 31.04\% & 9.227 & 43.73\% \\
        Q8 & 6.619 & 16.961 & 13.855 & 30.03\% & 12.760 & 40.62\% \\
        Q9 & 25.015 & 72.866 & 51.704 & 44.23\% & 47.161 & 53.72\% \\
        Q10 & 1.857 & 4.060 & 3.621 & 19.93\% & 3.554 & 22.95\% \\
        Q13 & 1.140 & 2.027 & 1.913 & 12.89\% & 1.870 & 17.77\% \\
        Q15 & 1.100 & 1.856 & 1.736 & 15.89\% & 1.701 & 20.48\% \\
        Q16 & 1.768 & 3.185 & 3.012 & 12.25\% & 2.992 & 13.61\% \\
        Q17 & 1.090 & 1.932 & 1.755 & 21.04\% & 1.748 & 21.88\% \\
        Q18 & 2.262 & 4.512 & 4.310 & 8.96\% & 4.245 & 11.87\% \\
        Q20 & 2.943 & 14.273 & 12.981 & 11.40\% & 8.035 & 55.06\% \\
        Q21 & 8.045 & 19.325 & 13.848 & 48.55\% & 13.643 & 50.37\% \\
        Q22 & 1.375 & 3.996 & 3.974 & 0.83\% & 3.963 & 1.25\% \\
        \hline
        \rowcolor{white}
        Total & 62.060 & 166.956 & 130.707 & 34.56\% & 118.529 & 46.17\% \\
        \hline
        \rowcolor{white}
        \\
        \multicolumn{7}{l}{I: Cost Based RW Cost OFF/Cache OFF/Educated Guess OFF (ms)} \\
        \multicolumn{7}{l}{II: Cost Based RW Cost ON/Cache OFF/Educated Guess OFF (ms)} \\
        \multicolumn{7}{l}{III: Cost Based RW Cost ON/Cache ON/Educated Guess OFF (ms)} \\
        \multicolumn{7}{l}{IV: III's improvement ratio} \\
        \multicolumn{7}{l}{V: Cost Based RW Cost ON/Cache ON/Educated Guess ON (ms)} \\
        \multicolumn{7}{l}{VI: V's improvement ratio} \\
    \end{tabular}
\end{table}

Of the 22 queries, 14 invoked at least one cost-based rewrite rule. We measured their compile times across four configurations.
The objective is to achieve query plans equivalent to those generated by the na\"ive cost-based method, but with significantly reduced optimization overhead. We measure its effectiveness by the relative reduction in overhead incurred by the na\"ive costing method, termed the \textbf{improvement ratio}.

Column I in \autoref{tab:tpch_table} shows the compilation time with cost disabled for rewrites. Column II shows the compilation time with the na\"ive method enabled. The compilation time increases by 269\%, almost 3 times, due to the na\"ive costing approach. 

The remaining columns quantify the performance gains from our contributions. Column III in \autoref{tab:tpch_table} lists the compilation time with only caching enabled. The overall improvement ratio of compilation time, as defined in \autoref{eq:IR2}, is 34.56\%, compared with the na\"ive method.

The effect of enabling both caching and early pruning using the educated-guess heuristic is shown in columns V and VI. The overall improvement ratio (\autoref{eq:IR1}) is 46.17\% relative to the na\"ive approach.

We validated that the query plans produced by the optimizer under both variants of our configuration—the proposed plan cache and the plan cache augmented with the educated guess—are identical to those generated by the baseline na\"ive method

The improvement ratio of the proposed framework over the na\"ive method is calculated as \autoref{eq:IR1}, with the corresponding results reported in Column VI. The last row of this column shows the overall result aggregated across all 14 queries:
\begin{equation}
ImprovementRatio_{VI} = \frac{column\:II - column\:V}{column\:II - column\:I} \label{eq:IR1}
\end{equation}
Here, the numerator represents the improvement achieved by our framework, while the denominator denotes the overhead introduced by the naïve costing method.

Similarly, the improvement ratio achieved using cached plans without the educated guess heuristic is defined by \autoref{eq:IR2}, with the results presented in column IV, the final row presents the overall result for the 14 queries:
\begin{equation}
ImprovementRatio_{IV} = \frac{column\:II -column\:III}{column\:II - column\:I} \label{eq:IR2}
\end{equation}

Across the 14 queries, the framework achieves an overall improvement ratio of \textbf{46.17\%}, indicating a substantial improvement of the compilation overhead from cost-based rewrite operations, while maintaining the same access plans as the na\"ive method.  Given that GaussDB currently supports a limited set of seven such rules, even greater performance gains are anticipated as the rule set expands.

The proposed method was further evaluated using the TPC-DS benchmark with a scale factor of 10 GB. All 99 queries comprising the benchmark workload were executed six times each, with reported results representing the average of the final three runs to capture steady-state performance characteristics. The underlying database tables were stored in row-oriented format. The experimental environment consisted of an x86 server equipped with 96 CPU cores and 768 GB of memory.

\autoref{tab:tpcds} details the TPC-DS queries that trigger cost-based rewrites in the GaussDB optimizer. Following the structure and notation of \autoref{tab:tpch_table} (for TPC-H), the table quantifies performance improvements relative to a na\"ive baseline. Column IV shows the improvement ratio when using cost estimation and caching exclusively (i.e., without the educated guess), while Column VI shows the improvement ratio with both caching and the educated guess enabled. The complete method yields a cumulative improvement ratio of \textbf{29.46\%} in planning time, without compromising plan quality—all generated plans remain equivalent to those produced by the na\"ive method.

\begin{sql*}
\begin{lstlisting}
SELECT a.No, a.Name, (SELECT Name FROM T1 WHERE CNo = a.CNo AND RTM = '00' FETCH FIRST 1 ROWS ONLY), 
  CodeName('relation',a.RTM) relation, CodeName('sex', a.Sex) sex, a.Birthday, 
  CodeName('idtype',a.IDType) idtype, a.IDNo, a.IDSDate, a.IDEDate, 
  (CASE a.IDLEF WHEN 'Y' THEN 'Yes' ELSE 'No' END) IDLEF, a.position, 
  (SELECT ocpname FROM T2 WHERE ocpcode=a.ocpcode), CodeName('ocptype',a.OcpType) ocptype, a.CPCode, 
  (SELECT codename FROM T3 WHERE codetype='nativeplace' AND code=a.nativeplace) nativeplace,
  (SELECT codename FROM T3 WHERE codetype='nativecity' AND code=a.nativecity) nativecity, 
  (SELECT SUM(Pm) FROM T4 WHERE CNo = a.CNo AND No = a.No) sumPm, 
  (SELECT SUM(c.SAPM) FROM T5 c WHERE c.cno = b.cno AND exists(SELECT 1 FROM T4 d  WHERE c.PolNo = d.PolNo AND c.CurPayToDate = d.PayToDate AND d.cno = b.cno AND d.no = a.No AND d.gcno = d.gcno union SELECT 1 FROM T6 d WHERE c.PolNo = d.PolNo AND c.CurPayToDate = d.PayToDate AND d.cno = b.cno AND d.no = a.No AND d.gcno = d.gcno )) SAPM, 
  NVL(TO_CHAR(( SELECT SUM(pam) FROM T7 WHERE otherno=a.gcno AND customerno=a.no AND customerflag='2')),'') sumpam,
  NVL(TO_CHAR(( SELECT SUM(pam) FROM T7 WHERE otherno=a.gcno AND customerno=a.no AND customerflag='1')),'') sumpam1, 
  NVL((SELECT CodeName('patype',lci.PaType) FROM T8 lci WHERE lci.GCNo=b.GCNo AND lci.No=a.No  AND lci.CPCode <> 'FM' ORDER BY makedate DESC FETCH FIRST 1 ROW ONLY), 
  (SELECT CodeName('patype',lbi.PaType) FROM T9 lbi WHERE lbi.GCNo=b.GCNo  AND lbi.No=a.No AND lbi.CPCode <> 'FM' ORDER BY makedate DESC  FETCH FIRST 1 ROW ONLY)) patype,
  (SELECT MIN(cvalidate) FROM T4 WHERE CNo = a.CNo AND No = a.No) cvalidate, CInvaliDate,
  (SELECT NVL(SUM(InsuAccBala), 0) FROM T10 WHERE CNo = a.CNo AND No = a.No) InsuAccBala, 
  (SELECT BankName FROM T11 WHERE BankCode = a.BankCode), a.BankCode, a.AccName, a.BankAccNo, b.CNo,
  (SELECT lbi.SchoolNmae FROM T9 lbi WHERE lbi.GCNo=b.GCNo AND lbi.No=a.No FETCH FIRST 1 ROWS ONLY),
  (SELECT lbi.ClASsName FROM T9 lbi WHERE lbi.GCNo=b.GCNo AND lbi.No=a.No  FETCH FIRST 1 ROWS ONLY),
  (SELECT CASE WHEN authorization_u = '1' THEN 'Y' ELSE 'No' END FROM T12 WHERE customerno = b.no) authorizationu, a.Remarks, 
  (SELECT lbi.EMail FROM T9 lbi WHERE lbi.GCNo=b.GCNo AND lbi.No=a.No FETCH FIRST 1 ROWS ONLY), 
  NVL((SELECT c.ENDorsementNo FROM T13 c WHERE c.gcno = a.gCNo AND c.no = a.no AND c.CNo = a.CNo AND c.FeeOperationType in ('NI','ZT','CT', 'XT', 'WT') ORDER BY ModifyDate,ModifyTime DESC FETCH FIRST 1 ROWS ONLY),'') ENDorsementNo, 
  NVL((SELECT (SELECT edorName FROM T14 WHERE edorCode = c.FeeOperationType AND appobj = 'G') FROM T13 c WHERE c.gcno = a.gCNo AND c.no = a.no AND c.CNo = a.CNo AND c.FeeOperationType in ('NI','ZT','CT', 'XT', 'WT') ORDER BY ModifyDate,ModifyTime DESC FETCH FIRST 1 ROWS ONLY),'') edorName, a.RuralPopulation, a.HANDicappedTypa,
  (SELECT CASE  WHEN AnamnesisFlag = 'Y' THEN 'pASt' WHEN AnamnesisFlag = 'N' THEN 'current' END FROM T9 WHERE GCNo = a.GCNo AND No = a.No FETCH FIRST 1 ROWS ONLY) AnamnesisFlag, 
  CASE WHEN c.individualbusiness = '1' THEN 'Yes' WHEN c.individualbusiness = '0' THEN 'No' END AS individualBusiness, 
  (SELECT CASE WHEN RuleAmntFlag = '1' THEN 'Yes' ELSE 'No' END FROM T15 WHERE prtno = a.prtno) AS RuleAmntFlag, 
  (SELECT SUM(amnt) FROM T4 WHERE cno = a.no) AS SumAmnt, (SELECT SUM(InsuAccBala)FROM T10 WHERE cno = a.cno) AS SumInsuAccBala, 
  (SELECT CASE  WHEN IntrductinoRiskFlag = '1' THEN 'Yes' WHEN IntrductinoRiskFlag = '0' THEN 'No' END FROM T16 WHERE CNo = a.CNo AND No = a.No ) IntrductinoRiskFlag, 
  (SELECT CASE  WHEN MarketResearchFlag = '1' THEN 'Yes' WHEN MarketResearchFlag = '0' THEN 'No' END FROM T16 WHERE CNo = a.CNo AND No = a.No ) MarketResearchFlag, 
  (SELECT CASE  WHEN InformationDateFlag = '1' THEN 'Yes' WHEN InformationDateFlag = '0' THEN 'No' END FROM T16 WHERE CNo = a.CNo AND No = a.No ) InformationDateFlag 
FROM T1 a LEFT JOIN T16 c ON c.no = a.no AND c.cno = a.cno, LCCont b 
WHERE a.CNo = b.CNo AND b.StateFlag = '1' AND a.GCNo = 'gcno' ORDER BY No LIMIT 1,10;
\end{lstlisting}
\caption{Customer example}
\label{sql:sqlexample3}
\end{sql*}

\begin{table}
    \centering
    \caption{TPC-DS Result}
    \label{tab:tpcds}
    \rowcolors{2}{gray!15}{white}
    \begin{tabular}{lrrrrrr}
        \hline
        No. & I & II & III & IV & V & VI \\
        \hline
        Q01 & 1.221 & 1.929 & 1.875 & 7.67\% & 1.874 & 7.81\% \\
        Q02 & 1.644 & 2.751 & 2.701 & 4.46\% & 2.757 & -0.60\% \\
        Q06 & 2.280 & 4.883 & 3.945 & 36.02\% & 3.917 & 37.10\% \\
        Q08 & 2.730 & 4.733 & 4.252 & 24.03\% & 4.229 & 25.15\% \\
        Q09 & 2.020 & 14.211 & 6.177 & 65.90\% & 3.118 & 91.00\% \\
        Q10 & 3.991 & 11.484 & 9.869 & 21.55\% & 9.907 & 21.05\% \\
        Q14 & 13.029 & 40.923 & 35.654 & 18.89\% & 36.008 & 17.62\% \\
        Q16 & 4.878 & 10.854 & 7.875 & 49.85\% & 7.565 & 55.03\% \\
        Q23 & 9.078 & 29.384 & 27.161 & 10.95\% & 22.033 & 36.20\% \\
        Q24 & 14.352 & 27.573 & 28.323 & -5.68\% & 28.243 & -5.07\% \\
        Q30 & 1.577 & 2.477 & 2.362 & 12.74\% & 2.332 & 16.11\% \\
        Q32 & 1.756 & 5.678 & 4.821 & 21.84\% & 4.797 & 22.46\% \\
        Q33 & 4.686 & 8.496 & 7.881 & 16.13\% & 7.863 & 16.60\% \\
        Q35 & 4.070 & 11.111 & 9.635 & 20.96\% & 9.703 & 20.00\% \\
        Q41 & 1.393 & 1.845 & 1.673 & 37.98\% & 1.410 & 96.17\% \\
        Q44 & 1.630 & 3.700 & 3.009 & 33.35\% & 3.059 & 30.97\% \\
        Q45 & 2.931 & 5.578 & 4.922 & 24.78\% & 4.880 & 26.37\% \\
        Q47 & 2.446 & 3.615 & 3.485 & 11.09\% & 3.410 & 17.54\% \\
        Q53 & 1.965 & 2.172 & 2.166 & 3.06\% & 2.182 & -4.83\% \\
        Q54 & 2.739 & 3.067 & 3.083 & -4.77\% & 3.087 & -5.88\% \\
        Q56 & 5.591 & 20.087 & 14.027 & 41.80\% & 14.005 & 41.96\% \\
        Q57 & 2.249 & 3.246 & 3.255 & -0.84\% & 3.159 & 8.76\% \\
        Q58 & 4.021 & 11.347 & 9.605 & 23.77\% & 9.672 & 22.86\% \\
        Q59 & 3.650 & 6.052 & 6.097 & -1.90\% & 5.969 & 3.44\% \\
        Q60 & 5.495 & 19.621 & 13.623 & 42.46\% & 13.601 & 42.62\% \\
        Q63 & 1.985 & 2.181 & 2.184 & -1.70\% & 2.193 & -5.94\% \\
        Q69 & 2.820 & 9.030 & 8.397 & 10.19\% & 8.413 & 9.93\% \\
        Q70 & 1.927 & 3.229 & 2.737 & 37.82\% & 2.736 & 37.87\% \\
        Q81 & 1.618 & 2.566 & 2.439 & 13.43\% & 2.420 & 15.47\% \\
        Q83 & 3.941 & 14.970 & 15.400 & -3.90\% & 15.311 & -3.10\% \\
        Q92 & 1.727 & 5.542 & 4.905 & 16.70\% & 4.836 & 18.51\% \\
        Q93 & 1.226 & 2.660 & 2.449 & 14.74\% & 2.670 & -0.67\% \\
        Q94 & 4.744 & 10.622 & 7.689 & 49.91\% & 7.397 & 54.87\% \\
        Q95 & 5.251 & 22.932 & 19.842 & 17.48\% & 15.729 & 40.74\% \\
        \hline
        Total & 126.659 & 330.548 & 283.520 & 23.07\% & 270.484 & 29.46\% \\
        \hline
        \rowcolor{white}
        \\
        \multicolumn{7}{l}{I: Cost Based RW Cost OFF/Cache OFF/Educated Guess OFF (ms)} \\
        \multicolumn{7}{l}{II: Cost Based RW Cost ON/Cache OFF/Educated Guess OFF (ms)} \\
        \multicolumn{7}{l}{III: Cost Based RW Cost ON/Cache ON/Educated Guess OFF (ms)} \\
        \multicolumn{7}{l}{IV: III's improvement ratio} \\
        \multicolumn{7}{l}{V: Cost Based RW Cost ON/Cache ON/Educated Guess ON (ms)} \\
        \multicolumn{7}{l}{VI: V's improvement ratio} \\
    \end{tabular}
\end{table}

Our implementation has been used in real-world customer benchmarks. In compliance with confidentiality agreements, the full benchmarks and associated queries cannot be disclosed. Nevertheless, we provide an illustrative example in \autoref{sql:sqlexample3} that demonstrates a complex, real-world workload successfully handled by our approach. To preserve confidentiality, table names and certain key column identifiers have been anonymized. For the query shown in \autoref{sql:sqlexample3}, plan generation without costing in QRW requires 13 ms. The baseline (na\"ive) method requires 205 ms, whereas our method produces an equivalent plan in 127 ms, representing a \textbf{59\%} improvement ratio.

Overall, our framework provides a more efficient method for applying cost-based rewrite rules and significantly lowers the barrier to supporting a greater number of rewrite rules in the optimizer. 

\section{Conclusion and Further Work}
\label{sec:futurework}
While many rewrite rules are universally beneficial, others require  cost-based selection. The computational overhead associated with this selection process is a major hurdle to adoption within the query optimizer. This paper proposes a framework for bottom-up query optimizers that leverages information available during the Query Rewrite (QRW) phase and fully exploits the potential of intermediate plan results by considering them at multiple levels of granularity. We further contribute methodological solutions for reusing the plan cache within a bottom-up optimizer architecture. Implemented in the GaussDB optimizer, this framework achieves an overall improvement ratio approximately \textbf{46.17\% on TPC-H} benchmark and \textbf{29.46\% on TPC-DS} benchmark compared to na\"ive methods. Although GaussDB currently implements only seven cost-based rules, the system is designed to accommodate a growing set. Thus, this work provides a scalable foundation for the future integration of more complex cost-based rewrite rules.

Our new framework offers some distinct advantages over other implementations: 

First, in contrast to the na\"ive method described in \autoref{sec:related}, our approach avoids the redundancy of regenerating execution plans from scratch for every cost-based rewrite rule evaluation. By strategically using \textbf{an educated guess} to establish a tight upper bound, we significantly prune the search space and accelerate the optimization process.

Second, unlike solutions that cache only at the query block level, our fine-grained intermediate caching is more robust. It enables reuse of cached intermediate plans even when rewrite rules alter the query block structure, ensuring high reuse rates and sustained efficiency gains throughout complex compilation sequences.

Finally, unlike the Mini CBO solution, our framework uses fully calculated costs, yielding more accurate plan cost estimations. This enhanced accuracy enables superior decision-making when applying rewrite rules, ensuring high plan quality without compromising optimization efficiency.

Although our current framework functions as a short-life cache for a single query compilation, the optimization data it gathers holds significant long-life potential. We plan to evolve this mechanism into a long-life cache that persists across compilation cycles, establishing it as a critical data source for advanced database functionalities such as:
\begin{itemize}
    \item \textbf{Materialized Query Tables (MQTs) and Statistics Views}: The cached intermediate plans can serve as a foundation for recommending or dynamically creating MQTs to accelerate query processing. Concurrently, they can help identify cardinality estimation errors, enabling the recommendation or creation of statistical views to correct them.
    \item \textbf{AI-Driven Optimizer Training}: The long-term accumulation of cached data provides an ideal training set for the next generation of AI-assisted query optimizers. This dataset offers real-world feedback on queries, accelerating the evolution of a more intelligent optimizer. Furthermore, we plan to leverage this data to train models specifically for the Educated Guess component of our framework.
\end{itemize}

\bibliographystyle{ACM-Reference-Format}
\bibliography{sample}

@String{Computing = "Computing" }

@String{Computer = "{IEEE} Computer" }

@inproceedings{Ahmed06,
author = {Ahmed, Rafi and Lee, Allison and Witkowski, Andrew and Das, Dinesh and Su, Hong and Zait, Mohamed and Cruanes, Thierry},
title = {Cost-based query transformation in Oracle},
year = {2006},
publisher = {VLDB Endowment},
booktitle = {Proceedings of the 32nd International Conference on Very Large Data Bases},
pages = {1026–1036},
numpages = {11},
location = {Seoul, Korea},
series = {VLDB '06}
}

@online{baba,
  title         = "Query rewrites",
  url           = "https://www.alibabacloud.com/help/en/polardb/polardb-for-mysql/user-guide/query-rewrite",
  organization = {Alibaba Group},
  author = "Alibaba Cloud",
  year = "2025",
  urldate = "2025-12-12"
}

@inproceedings{Pirahesh92,
author = {Pirahesh, Hamid and Hellerstein, Joseph M. and Hasan, Waqar},
title = {Extensible/rule based query rewrite optimization in Starburst},
year = {1992},
isbn = {0897915216},
publisher = {Association for Computing Machinery},
address = {New York, NY, USA},
url = {https://doi.org/10.1145/130283.130294},
doi = {10.1145/130283.130294},
booktitle = {Proceedings of the 1992 ACM SIGMOD International Conference on Management of Data},
pages = {39–48},
numpages = {10},
location = {San Diego, California, USA},
series = {SIGMOD '92}
}

@techreport{Chaudhuri99,
  author      = {Chaudhuri, Surajit},
  title       = {Accurate Query Optimization by Sub-plan Memoization},
  institution = {Microsoft Research},
  year        = {1999},
  number      = {MSR-TR-99-87},
  url         = {https://www.microsoft.com/en-us/research/publication/accurate-query-optimization-by-sub-plan-memoization/}
}

@article{Ding24,
author = {Ding, Bailu and Narasayya, Vivek and Chaudhuri, Surajit},
title = {Extensible Query Optimizers in Practice},
year = {2024},
issue_date = {Dec 2024},
publisher = {Now Publishers Inc.},
address = {Hanover, MA, USA},
volume = {14},
number = {3–4},
issn = {1931-7883},
url = {https://doi.org/10.1561/1900000077},
doi = {10.1561/1900000077},
journal = {Found. Trends Databases},
month = dec,
pages = {186–402},
numpages = {219}
}

@INPROCEEDINGS{Pirahesh97,
  author={Pirahesh, H. and Leung, T.Y.C. and Hasan, W.},
  booktitle={Proceedings 13th International Conference on Data Engineering}, 
  title={A rule engine for query transformation in Starburst and IBM DB2 C/S DBMS}, 
  year={1997},
  volume={},
  number={},
  pages={391-400},
  doi={10.1109/ICDE.1997.581945}}

@inproceedings{Yan95,
author = {Yan, Weipeng P. and Larson, Per-\r{A}ke},
title = {Eager Aggregation and Lazy Aggregation},
year = {1995},
isbn = {1558603794},
publisher = {Morgan Kaufmann Publishers Inc.},
address = {San Francisco, CA, USA},
booktitle = {Proceedings of the 21th International Conference on Very Large Data Bases},
pages = {345–357},
numpages = {13},
series = {VLDB '95}
}

@inproceedings{Flajolet07,
  author    = {Flajolet, Philippe and Fusy, {\'E}ric and Gandouet, Olivier and Meunier, Fr{\'e}d{\'e}ric},
  title     = {{HyperLogLog}: The Analysis of a Near-Optimal Cardinality Estimation Algorithm},
  booktitle = {Proceedings of the 2007 Conference on Analysis of Algorithms (AofA '07)},
  year      = {2007},
  pages     = {137--156},
  publisher = {Discrete Mathematics and Theoretical Computer Science},
}

@inproceedings{Zeyl25,
author = {Zeyl, Tim and Cheng, Qi and Pournaghi, Reza and Lam, Jason and Wang, Weicheng and Wong, Calvin and Chen, Chong and Larson, Per-Ake},
title = {Including Bloom Filters in Bottom-up Optimization},
year = {2025},
isbn = {9798400715648},
publisher = {Association for Computing Machinery},
address = {New York, NY, USA},
url = {https://doi.org/10.1145/3722212.3724440},
doi = {10.1145/3722212.3724440},
booktitle = {Companion of the 2025 International Conference on Management of Data},
pages = {703–715},
numpages = {13},
location = {Berlin, Germany},
series = {SIGMOD/PODS '25}
}

@article{Zhu05,
  author    = {Zhu, Qiang and Tao, Yingying and Zuzarte, Calisto},
  title     = {Optimizing complex queries based on similarities of subqueries},
  journal   = {Knowledge and Information Systems},
  year      = {2005},
  volume    = {8},
  number    = {3},
  pages     = {350--373},
  doi       = {10.1007/s10115-004-0189-y}
}

@inproceedings{Meduri11,
author = {Vamsikrishna, Meduri Venkata and Tan, Kian-Lee},
title = {Subquery plan reuse based query optimization},
year = {2011},
publisher = {Computer Society of India},
address = {Mumbai, Maharashtra, IND},
booktitle = {Proceedings of the 17th International Conference on Management of Data},
articleno = {11},
numpages = {12},
location = {Bangalore, India},
series = {COMAD '11}
}

@misc{Song25,
      title={QUITE: A Query Rewrite System Beyond Rules with LLM Agents}, 
      author={Yuyang Song and Hanxu Yan and Jiale Lao and Yibo Wang and Yufei Li and Yuanchun Zhou and Jianguo Wang and Mingjie Tang},
      year={2026},
      eprint={2506.07675},
      archivePrefix={arXiv},
      primaryClass={cs.DB},
      url={https://arxiv.org/abs/2506.07675}, 
}

@article{Sun2024RBotCoRR,
author = {Sun, Zhaoyan and Zhou, Xuanhe and Li, Guoliang and Yu, Xiang and Feng, Jianhua and Zhang, Yong},
title = {R-Bot: An LLM-Based Query Rewrite System},
year = {2025},
issue_date = {August 2025},
publisher = {VLDB Endowment},
volume = {18},
number = {12},
issn = {2150-8097},
url = {https://doi.org/10.14778/3750601.3750625},
doi = {10.14778/3750601.3750625},
journal = {Proc. VLDB Endow.},
month = aug,
pages = {5031–5044},
numpages = {14}
}

@misc{Liu2024Query,
      title={GenRewrite: Query Rewriting via Large Language Models}, 
      author={Jie Liu and Barzan Mozafari},
      year={2025},
      eprint={2403.09060},
      archivePrefix={arXiv},
      primaryClass={cs.DB},
      url={https://arxiv.org/abs/2403.09060}, 
}

@inproceedings{Zhou2007Efficient,
author = {Zhou, Jingren and Larson, Per-Ake and Freytag, Johann-Christoph and Lehner, Wolfgang},
title = {Efficient exploitation of similar subexpressions for query processing},
year = {2007},
isbn = {9781595936868},
publisher = {Association for Computing Machinery},
address = {New York, NY, USA},
url = {https://doi.org/10.1145/1247480.1247540},
doi = {10.1145/1247480.1247540},
booktitle = {Proceedings of the 2007 ACM SIGMOD International Conference on Management of Data},
pages = {533–544},
numpages = {12},
location = {Beijing, China},
series = {SIGMOD '07}
}

@article{Graefe1995Cascades,
  author    = {Goetz Graefe},
  title     = {The Cascades Framework for Query Optimization},
  journal   = {IEEE Data Eng. Bull.},
  volume    = {18},
  number    = {3},
  year      = {1995},
  pages     = {19--29},
  ee        = {db/journals/debu/Graefe95a.html},
  bibsource = {DBLP, http://dblp.uni-trier.de}
}

@INPROCEEDINGS{Graefe1991Volcano,
  author={Graefe, G. and McKenna, W.J.},
  booktitle={Proceedings of IEEE 9th International Conference on Data Engineering}, 
  title={The Volcano optimizer generator: extensibility and efficient search}, 
  year={1993},
  volume={},
  number={},
  pages={209-218},
  doi={10.1109/ICDE.1993.344061}}

@inproceedings{Soliman2014Orca,
author = {Soliman, Mohamed A. and Antova, Lyublena and Raghavan, Venkatesh and El-Helw, Amr and Gu, Zhongxian and Shen, Entong and Caragea, George C. and Garcia-Alvarado, Carlos and Rahman, Foyzur and Petropoulos, Michalis and Waas, Florian and Narayanan, Sivaramakrishnan and Krikellas, Konstantinos and Baldwin, Rhonda},
title = {Orca: a modular query optimizer architecture for big data},
year = {2014},
isbn = {9781450323765},
publisher = {Association for Computing Machinery},
address = {New York, NY, USA},
url = {https://doi.org/10.1145/2588555.2595637},
doi = {10.1145/2588555.2595637},
booktitle = {Proceedings of the 2014 ACM SIGMOD International Conference on Management of Data},
pages = {337–348},
numpages = {12},
location = {Snowbird, Utah, USA},
series = {SIGMOD '14}
}

@article{GaussDB2020,
author = {Li, Guoliang and Tian, Wengang and Zhang, Jinyu and Grosman, Ronen and Liu, Zongchao and Li, Sihao},
title = {GaussDB: A Cloud-Native Multi-Primary Database with Compute-Memory-Storage Disaggregation},
year = {2024},
issue_date = {August 2024},
publisher = {VLDB Endowment},
volume = {17},
number = {12},
issn = {2150-8097},
url = {https://doi.org/10.14778/3685800.3685806},
doi = {10.14778/3685800.3685806},
journal = {Proc. VLDB Endow.},
month = aug,
pages = {3786–3798},
numpages = {13}
}

@INPROCEEDINGS{Memarzia2024GaussDB,
  author={Memarzia, Puya and Zhang, Huaxin and Ho, Kelvin and Grosman, Ronen and Wang, Jiang},
  booktitle={2024 IEEE 40th International Conference on Data Engineering (ICDE)}, 
  title={GaussDB-Global: A Geographically Distributed Database System}, 
  year={2024},
  volume={},
  number={},
  pages={5111-5118},
  doi={10.1109/ICDE60146.2024.00383}}

@inproceedings{Wang2022WeTune,
author = {Wang, Zhaoguo and Zhou, Zhou and Yang, Yicun and Ding, Haoran and Hu, Gansen and Ding, Ding and Tang, Chuzhe and Chen, Haibo and Li, Jinyang},
title = {WeTune: Automatic Discovery and Verification of Query Rewrite Rules},
year = {2022},
isbn = {9781450392495},
publisher = {Association for Computing Machinery},
address = {New York, NY, USA},
url = {https://doi.org/10.1145/3514221.3526125},
doi = {10.1145/3514221.3526125},
booktitle = {Proceedings of the 2022 International Conference on Management of Data},
pages = {94–107},
numpages = {14},
location = {Philadelphia, PA, USA},
series = {SIGMOD '22}
}

@article{Li2024LLMR2CoRR,
author = {Li, Zhaodonghui and Yuan, Haitao and Wang, Huiming and Cong, Gao and Bing, Lidong},
title = {LLM-R2: A Large Language Model Enhanced Rule-Based Rewrite System for Boosting Query Efficiency},
year = {2024},
issue_date = {September 2024},
publisher = {VLDB Endowment},
volume = {18},
number = {1},
issn = {2150-8097},
url = {https://doi.org/10.14778/3696435.3696440},
doi = {10.14778/3696435.3696440},
journal = {Proc. VLDB Endow.},
month = sep,
pages = {53–65},
numpages = {13}
}

@misc{Dharwada2026LITHE,
      title={Query Rewriting via LLMs}, 
      author={Sriram Dharwada and Himanshu Devrani and Jayant Haritsa and Harish Doraiswamy},
      year={2025},
      eprint={2502.12918},
      archivePrefix={arXiv},
      primaryClass={cs.DB},
      url={https://arxiv.org/abs/2502.12918}, 
}

@article{Zhou2021Learned,
author = {Zhou, Xuanhe and Li, Guoliang and Chai, Chengliang and Feng, Jianhua},
title = {A learned query rewrite system using Monte Carlo tree search},
year = {2021},
issue_date = {September 2021},
publisher = {VLDB Endowment},
volume = {15},
number = {1},
issn = {2150-8097},
url = {https://doi.org/10.14778/3485450.3485456},
doi = {10.14778/3485450.3485456},
journal = {Proc. VLDB Endow.},
month = sep,
pages = {46–58},
numpages = {13}
}

\end{document}